\documentclass[letterpaper,twocolumn,10pt]{article}
\usepackage{usenix,epsfig,url, subfigure,verbatim}
\begin{document}
\pdfoutput=1

\newcommand{\woj}[1]{\ \\ \fbox{\parbox{1.0\linewidth}{\bf {\sc Wojciech}: #1}}}
\newcommand{\aaa}[1]{\ \\ \fbox{\parbox{1.0\linewidth}{\bf {\sc Alvin}: #1}}}
\newcommand{\remove}[1]{}

\date{}

\title{\Large \bf Toward a Principled Framework for Benchmarking Consistency}

\author{
{\rm Muntasir Raihan Rahman}\\
HP Labs, Palo Alto / \\
University of Illinois at Urbana Champaign \\
{\rm \url{mrahman2@illinois.edu}}
\and
{\rm Wojciech Golab, Alvin AuYoung} \\
{\rm Kimberly Keeton, Jay J.\ Wylie}\\
HP Labs, Palo Alto \\
{\rm \url{firstname.lastname@hp.com}}
} 

\maketitle

\thispagestyle{empty}


\subsection*{Abstract}

Large-scale key-value storage systems sacrifice consistency in the
interest of dependability (i.e., partition-tolerance and
availability), as well as performance (i.e., latency).  Such systems
provide eventual consistency, which---to this point---has been difficult to quantify in real systems.
Given the many implementations and deployments of
eventually-consistent systems (e.g., NoSQL systems), attempts have
been made to measure this consistency empirically, but they suffer
from important drawbacks.  For example, state-of-the art consistency
benchmarks exercise the system only in restricted ways and disrupt the
workload, which limits their accuracy.

In this paper, we take the position that a consistency benchmark
should paint a comprehensive picture of the relationship between the
storage system under consideration, the workload, the pattern of
failures, and the consistency observed by clients.  To illustrate our
point, we first survey prior efforts to quantify eventual consistency.
We then present a benchmarking technique that overcomes the shortcomings of
existing techniques to measure the consistency observed by clients as 
they execute the workload under consideration.  This method is 
versatile and minimally disruptive to the system under test.  As a
proof of concept, we demonstrate this tool on Cassandra.



\section{Introduction}

Large-scale key-value storage systems are quickly becoming an
essential component of many IT infrastructures. From fast-growing
start-ups to large enterprises, these systems are
becoming commonplace in production use because of their ability to
scale easily and the availability of many widely-supported software
implementations.  However, in order to provide performance and
dependability at scale, the common principle followed by these
key-value systems is to relax data consistency \cite{Vogels}.

As these systems find their way into a wider variety of industries, it
becomes increasingly important to understand the implications of this
relaxed consistency model: to what extent relaxation improves
system performance and to what extent it degrades data consistency.

For example, Web-based applications rely on key-value systems to
provide high-throughput and low-latency access to content. While these
applications do not strictly require serializability
for correct operation, they may require a stronger
property than eventual consistency, such as causal or ``causal+''
\cite{cops:sosp11} consistency, in order to improve the user
experience.



On the other hand, cloud-based health care applications likely value
predictable consistency over performance.  Eventually consistent
updates to a patient's record may introduce mistakes along the path of
patient care.  For example, stale information (e.g., due to weak
consistency) about a patient's dosage or medical history may lead to
incorrect, or---in an extreme case---harmful treatment plans.

Today, cloud customers who care about consistency have limited means
to understand or control data consistency when choosing among
available storage systems, or their configurations. For example,
decisions to tune ``knobs'' such as the replication factor or
quorum size remain ad-hoc, and may lead to excessive replication or
operational costs.  More importantly, no combination of these
knob settings can ensure that the storage system is strongly consistent (e.g.,
always returns the freshest data).  This shortcoming is a fundamental
limitation of such always-available, partition-tolerant systems, as
observed by Brewer \cite{brew:cap} and formalized by Lynch et
al.\ \cite{lg:cap}. Moreover, many modern systems often choose to
further sacrifice consistency for better performance \cite{abadi:cap}.

We argue that a methodology for comprehensive consistency measurement
is necessary to evaluate today's eventually consistent systems.  Such
a measurement framework can identify the shortcomings of architectural
designs or implementation errors in existing systems. Moreover, it can
determine the actual consistency behavior of a particular deployment,
which may be helpful to guide configuration and deployment decisions.


Prior techniques for measuring consistency follow a methodology that
is oversimplified, and as a result suffer from important drawbacks.
For example, the act of measurement disrupts the workload by injecting
operations, causing a troublesome ``observer effect''.  Moreover, the
injected operations tend to stress the system, which may elicit
worst-case behavior even for a light workload. Understanding 
observed, as opposed to worst-case, consistency is important for
systems designers considering performance trade-offs, particularly if
observed consistency is vastly different from the worst-case.

Our position is simple---a consistency benchmark should
produce precise and accurate measurements of consistency with minimal
disruption to the system under evaluation.  These measurements should
reflect the consistency actually observed by clients in the workload
under consideration, rather than the consistency of operations
injected artificially into the workload.  
Furthermore, a benchmark must collect measurements in a system-agnostic way,
enabling comparisons not only between different implementations of the
\emph{same} consistency model (e.g., sloppy quorums \cite{vogels:dynamo}),
but also between different consistency models.

In this paper, we describe a principled approach to consistency
measurement that captures more faithfully and accurately the
actual consistency behavior of a key-value storage system
for an arbitrary workload.
Our specific contributions are:
\vspace{-2.5mm}
\begin{enumerate}\setlength{\itemsep}{-1.5mm}
\item A survey of known techniques for quantifying and benchmarking consistency, and discussion of their limitations (Section~\ref{sec:rel}).
\item An outline of a more general and precise approach to consistency measurement (Section~\ref{sec:pos}).
\item A proof-of-concept benchmarking tool, which we use to obtain consistency measurements for the Cassandra \cite{cassie} key-value store (Section~\ref{sec:exp}).
\end{enumerate}




\section{Related work} \label{sec:rel}
Consistency in this paper refers to the notion that different clients
accessing a storage system agree in some way on the state of data.  In
the literature, this is termed the \emph{client-centric}
view, as opposed to the \emph{data-centric} view, which refers to
details that are not directly observable by clients (e.g., messages in
flight, state of replicas).  The client-centric view is more natural
in the context of benchmarking consistency, as it does not require
system-specific and disruptive instrumentation to collect intimate details
of the execution.
Instead, it considers only the information that clients can capture locally
as they apply \emph{get} and \emph{put} operations on keys, such
as the start and end time of each operation as well as its arguments
and response.

Client-centric definitions of consistency typically refer to agreement
on when and in which order operations take effect (e.g., see \cite{terry:baseball}).
As we discuss shortly, early attempts to benchmark consistency
focus on the ``when'', and interpret this question as meaning roughly
``How soon after a write operation returns do read operations
return the written value?'', or in other words, ``How eventual is 
eventual?'' \cite{Bermbach,ycsbpp,wada:cidr}.


Formalizing and answering these questions precisely bring us a step
closer to understanding the complex relationship between
the workload applied to a storage system, the failure pattern,
the configuration parameters, and the observed client-centric consistency.
In contrast, prior work covers a narrow sub-space of this multidimensional relationship
that considers only failure-free executions,
and relies on an informal methodology that exercises the storage system
only near the limits of its ``consistency envelope''.

\remove{  
For example, if a data item is in
the same state for some period of time (i.e., is not being written),
then a system where all the read operations return the latest value of
the data item is more consistent than one where a read returns a stale
or incorrect value.  Or, if the state of the data item is changing,
then a system where different clients observe the sequence of state
changes in the same order is more consistent than one where two
clients observe the sequence of states in a different orders.
}

\begin{figure}[!t]
	\centering	
	\subfigure{%
		\includegraphics[width=\linewidth]{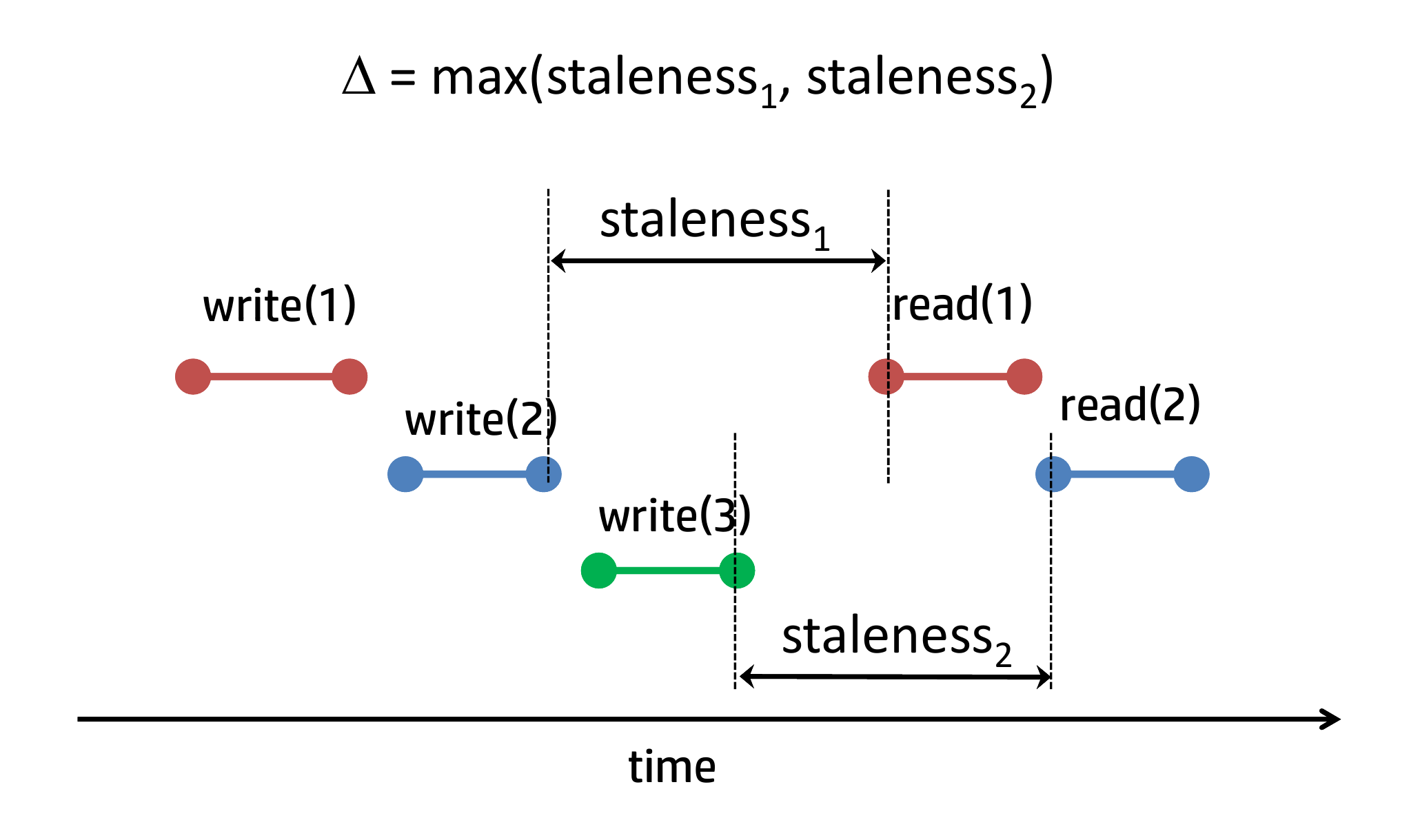}%
	}
	\caption{Example calculation of $\Delta$.}
	\label{fig:delta-extended}
	\vspace{-10pt}
\end{figure}

\vspace{-6pt}
\paragraph{Definitions of version and time-based staleness}\ \\
Staleness is a fundamental concern in data management, and can be used
to describe the quality of both the data and the system that stores
it.  In this benchmark, we focus on the quality of the storage system,
and in particular the protocol synchronizing different replicas of
data.  To that end, we consider staleness as a relative measure: how
long ago was the value read first updated (e.g.,
see Figure~\ref{fig:delta-extended}).  In other words, data becomes
stale the first time it is overwritten by newer data.

Prior techniques for quantifying staleness in key-value storage
systems either count versions (e.g., the value read is the
second-latest value written) or measure time (e.g., the value read
is one hour older than the latest value written) \cite{aiyer, gls:fun,
  ksc:qos, yv:costs, conit, zz:trading}.
These quantities are easy to state precisely under the simplifying
assumption that read and write operations are instantaneous---a
collection of unique points on a one-dimensional axis.  In that case,
there is a natural total order on the operations, and moreover the
``latest value'' at any point in time is well defined.  In contrast,
in real-world scalable storage systems, operational latencies due to
processing, networking and I/O are non-trivial, and so there can be multiple
operations in flight at any given time, even on a single key.  Thus,
non-trivial latencies and parallelism complicate reasoning
about when a given operation takes effect, as well as the order in
which operations take effect relative to each other.

A more precise treatment of staleness devised by the theory community
includes the (client-centric) concepts of $k$-atomicity \cite{aiyer} and $\Delta$-atomicity
\cite{gls:fun}.  The $k$-atomicity property is a natural formalization
of version-based staleness.  An execution of operations in a key-value store is
$k$-atomic if the operations in that execution can be totally ordered
so that: (1) the total order extends the ``happens before''
partial order (i.e., if operation $A$ ended before operation $B$ began
during the execution, then $A$ precedes $B$ in the total order); and
(2) each read returns the value assigned by one of the $k$ most recent
writes preceding the read in the total order.  (In the case $k=1$,
$k$-atomicity corresponds to Lamport's atomicity concept
\cite{lam:ipc}, which we discuss below.)  For any given execution,
we can quantify version-based staleness by solving the following
optimization problem: find the smallest $k$ for which the execution is
$k$-atomic.  We are not aware of an efficient (i.e., poly-time)
solution to this problem, although \cite{gls:fun} presents progress
toward solving the corresponding decision problem for $k=2$.

The $\Delta$-atomicity property attempts to capture time-based
staleness by stating that read operations must return values that are
at most $\Delta$ time units staler than the latest value for a key.
More formally, if we ``stretch'' the start time of each read to a
point $\Delta$ time units earlier, then the resulting execution should
be atomic in Lamport's sense \cite{lam:ipc}.  For any given
execution, it is possible to compute the smallest $\Delta \geq 0$ for
which that execution is $\Delta$-atomic using an efficient algorithm
\cite{gls:fun}.

Figure~\ref{fig:delta-extended} illustrates $\Delta$ in action.  The
start and end times are shown for three writes and two reads, all
operating on the same key.  We assume that each operation
takes effect between its beginning and end.  For example, 2 is the latest value from
the moment \textsf{write(2)} ends, and possibly even earlier.  Thus,
\textsf{read(1)} returns a value that is stale by at least the width
of the ``gap'' between it and \textsf{write(2)}. Even though
\textsf{write(3)} is the latest value, staleness for \textsf{read(1)} is measured from
the \emph{first} unseen update to the key: \textsf{write(2)}.  Similarly,
the staleness for \textsf{read(2)} is measured from the end of 
\textsf{write(3)}.

For completeness, we also briefly discuss well-studied notions of weakly
consistent shared objects from distributed computing theory
literature.  Lamport proposed the notions of \emph{safe},
\emph{regular} and \emph{atomic registers} (i.e., shared objects that
support read and write operations).  These specifications describe the
correct behavior of read operations when they can execute concurrently
with writes and with each other, but do not adequately capture the
possibility that non-concurrent operations may appear to take effect out of
order---a commonplace phenomenon in modern quorum-replicated systems.
Lamport's atomicity property is similar in spirit to Herlihy and Wing's
\emph{linearizability} \cite{her:lin} and Papadimitriou's \emph{strict
  serializability} \cite{papa:ss} for read/write register objects.

\vspace{-6pt}
\paragraph{Measuring and bounding staleness}\ \\
Several papers attempt to measure or bound staleness in order to
characterize the spectrum of trade-offs surrounding Brewer's celebrated
CAP principle \cite{abadi:cap, brew:cap}.  
Wada et al.\ \cite{wada:cidr} measure time-based staleness in cloud
storage platforms by writing timestamps to a key from one client three
times per second, reading the same key from another client fifty times
per second, and computing the difference between the reader's local
time and the timestamp read.  In experiments using
Amazon's SimpleDB \cite{simpledb}, they observe staleness on
the order of seconds.

The methodology of Wada et al.\ is sufficient to obtain evidence
relevant to their central research question---whether cloud storage
systems in practice provide more consistency than they promise.
However, their technique also has several disadvantages as a
consequence of exercising the system in an artificial way.  First, the
measurement is disruptive because it introduces additional write
operations to the workload.  This is unsuitable in a production
environment, unless the operations are applied to a special ``dummy''
key, in which case the outcome may not predict accurately the
staleness observed by reads on the other keys.  Secondly, the
technique is pessimistic because it considers a pattern of access
where read operations occur back-to-back with writes.  This
measurement captures the minimum time needed for replicas of a
key-value pair to synchronize, but in a real world workload, gaps
between operations may result in clients observing far less staleness.
In particular, if the load is trivial then it is possible that all
operations (except the ones injected artificially) will be atomic.  A
third drawback is the use of only a single writer.  While this
certainly simplifies calculations, the measurements obtained may fail
to cover special execution paths of the storage system for dealing
with concurrent writes, which hurts accuracy further.
  
Bermbach et al.\ \cite{Bermbach} and Patil et al.\ \cite{ycsbpp}
measure staleness using techniques similar to Wada et al.  The
latter paper presents an extension of the Yahoo Cloud Serving
Benchmark (YCSB) \cite{ycsb}, with support for basic consistency
benchmarking.  Their technique relies on a middleware service,
namely ZooKeeper \cite{zoo},
to convey timing information between readers and writers.
This technique is limited in precision due to the latency
introduced by operations on ZooKeeper, and hence it produces
results with one-sided error:
reported consistency violations are true assuming synchronized
clocks, but lack of reported violations does not imply
atomic behavior.

Bailis et al.\ \cite{pbs} consider the problem of predicting the
staleness from an abstract model of the storage system, including
details such as the distribution of latencies for network links.  This
work considers both version and time-based staleness, and provides an
upper bound on the probability that a client observes stale data.
This prediction, similar to the measurements of Wada et al., may be
overly pessimistic for light workloads.  Predicting and measuring
staleness are complementary techniques---prediction can be used for
planning and measurement can be used in a variety of ways, such as
performance tuning, monitoring, evaluating service-level agreements,
and feedback control.

\vspace{-6pt}
\paragraph{Other work}\ \\
Shapiro et al.\ formalize eventual consistency for
shared objects that avoid conflicts by design, for example
by providing commutative operations \cite{cfrdt, ccrd}.
Conventional key-value storage systems, like Cassandra, fall
outside this category because write operations are inherently conflict-prone.
Zhu et al.\ \cite{zhu} give formal definitions of eventual
consistency for read/write storage systems,
as well as several client-centric properties:
read-your-writes, monotonic reads, writes follow reads, and monotonic
writes.  This work does not provide a way to measure the difference
between a particular consistency property and the actual consistency
delivered by a storage system.
Less formal definitions of eventual consistency appear in numerous papers
(e.g., \cite{bayou,Vogels}).

\remove{
\begin{enumerate}
\item CAP-type trade-offs \cite{brew:cap,lg:cap,abadi:cap}.
\item Reference \cite{zhu} -- They give formal client-centric definitions of read-your-writes, monotonic reads, writes follow reads,
      and monotonic writes.  They also define eventual consistency, but that definition is not client-centric.  The paper is theoretical and has no empirical component. In this paper ``verification'' means ``proof sketch''
\item Prince Mahajan convergence paper
\begin{itemize}
\item eventually consistent systems provide more than they promise in practice
\item when a system allows the user to choose consistency settings, consistency with stronger settings was not much better than with weak settings
\end{itemize}
\item YCSB / YCSB++
\begin{itemize}
\item intrusive and not very precise due to the use of ZooKeeper
\item single-writer only for the key being tested
\end{itemize}
\item HP Labs PODC paper
\begin{itemize}
\item looks at verification algorithms only, no empirical evaluation
\item lacks the concept of ``instantaneous consistency''
\end{itemize}
\item PBS -- Peter Bailis et al.
\begin{itemize}
\item gives a worst-case probabilistic bound on staleness, but how tight are the bounds?
\item prediction, not measurement
\end{itemize}
\item consistency rationing -- VLDB 2012
\begin{itemize}
\item talks about price more than CAP-type trade-offs and benchmarking
\end{itemize}
\item Epsilon-serializability \cite{epsilon} -- not looking at staleness
\end{enumerate}
}



\section{Toward a benchmarking framework} \label{sec:pos}

We focus on creating a client-centric benchmarking tool that 
measures observed consistency and is
minimally disruptive to the system under evaluation.  Since
consistency and fault-tolerance are intimately related in eventually
consistent systems, the tool should provide support for fault
injection.  This includes crashes (individual and correlated) as well
as network partitions, and necessitates ``white-box'' access to the
infrastructure.  Finally, the tool should simplify analysis of the
results by presenting useful visualizations to the user.

As a stepping stone towards building a comprehensive benchmarking
framework, we now describe a methodology for minimally disruptive
measurement of consistency in arbitrary workloads.  We then suggest
how such measurements might be visualized.  Since our methodology is
client-centric, it can be married with any workload generator.  The
measurement entails collecting timing information at clients for an
arbitrary interleaving of operations, and calculating consistency
metrics only from this information using theoretically-sound
techniques.  As a running example, we consider the calculation of
the $\Delta$ quantity described in Section~\ref{sec:rel}, and then
discuss integration with YCSB~\cite{ycsb}.

$\Delta$-atomicity is defined abstractly for arbitrary
executions, including ones containing concurrent writes to the same
key.  To quantify staleness, we propose to calculate $\Delta$ for a
given execution using the procedure described in \cite{gls:fun}.
First, we group operations into \emph{clusters}---sets of operations
that access the same key and read or write same value \cite{gktest}.
For example, in Figure~\ref{fig:delta-extended} there are three
clusters, red, blue and green, corresponding to the values 1, 2 and 3.
Next, we choose a key $k$ and for each pair of clusters for that key,
and we determine the staleness due to the interaction of operations in
these clusters by evaluating a \emph{scoring function $\chi$}
\cite{gls:fun}.  We omit the formal details and point out only that in
Figure~\ref{fig:delta-extended}, $\chi$ is the width of the staleness ``gaps''
experienced by \textsf{read(1)} and \textsf{read(2)}.  Finally, we compute
the $\Delta$ value for key $k$ by taking the maximum of $\chi$ over
all pairs of clusters for $k$.  We repeat for each key and, taking the
maximum, obtain a global $\Delta$ indicating the staleness for the
entire execution.  Note that since the calculation combines time
values from multiple hosts, accuracy is contingent upon synchronized
clocks.

The quantities $\chi$ and $\Delta$ can be displayed visually in
various ways.  For example, using $\Delta$'s for different keys, we
can plot a histogram that shows what proportion of the key space was
read in a consistent manner.  Or, using $\chi$ values for one key, we
can plot a histogram that shows what proportion of clusters contained
reads of stale values (which, in turn, estimates what proportion of
reads returned stale values).  We can also use a time series plot of 
$\chi$ to visualize the \emph{instantaneous
  consistency} in an execution, which indicates the staleness of
values read at different points in time.  This allows us to observe
how staleness varies over time (e.g., in response to load spikes
or failures), information that is masked by $\Delta$ alone since it
quantifies consistency for the duration of an entire execution.  Note
that $\chi$ and $\Delta$, as well as the corresponding visualizations,
can be obtained for a subset of the key space (e.g., chosen
through random sampling).




\section{Experimental evaluation} \label{sec:exp}

To demonstrate our benchmarking methodology, we integrated our consistency measurement
  technique into YCSB \cite{ycsb},
  and used the modified YCSB to measure consistency in Cassandra \cite{cassie},
  a widely adopted key-value storage system.
Our experiments use YCSB v.\ 0.1.4 and Cassandra v.\ 1.1.0.

The experimental hardware platform is a cluster of ten commodity
dual-socket 6-core Xeon servers equipped with 1GigE network interface
cards and 96GB DRAM.  Each server ran a 32-thread YCSB client on one
socket, and a Cassandra node on the other socket, configured with
default options except as follows: keys were hashed uniformly across
all nodes and 3-way replicated using the ``simple'' replica placement
strategy \cite{cassiewiki}. By default, the Cassandra connector in
YCSB used consistency level ``ONE'' for both reads and writes. This
consistency level requires that a write be applied to the commit log
and memory table of at least one replica node before returning to the
client, and allows a read to return the value obtained from the first
replica that responds.

We instrumented the YCSB source code to log timing information for
each operation using a millisecond-precision clock.  We pre-loaded
Cassandra with 1000 keys and applied a read-heavy (80\% get, 20\%
put) workload for 60 seconds.  The keys were drawn from YCSB's ``hot
spot'' distribution, with 80\% of the operations going to a subset of
hot keys comprising 20\% of the key space.

\remove{
Using the collected timing information, we compute staleness using the
$\Delta$ quantity defined in \cite{gls:fun}.  First, we group
operations into \emph{clusters}---groups of operations that access
(i.e., read or write) the same value \cite{gktest}.  Next, we determine conflicts
between pairs of clusters by evaluating a \emph{scoring function $\chi$},
which is defined formally in \cite{gls:fun}.  At a high level, the
scoring function quantifies the staleness of data due to consistency
violations between operations in two clusters, and indicates the
relative staleness observed by read operations. It is measured in
units of time and has a range from zero to infinity.  The maximum of
the score function for any pair of clusters is the $\Delta$-value for
the execution, which indicates the worst-case relative staleness
observed in the workload.
}

\begin{figure}[tbp]
	\centering
	\vspace{-10pt}	
	\subfigure{%
		\includegraphics[width=\linewidth]{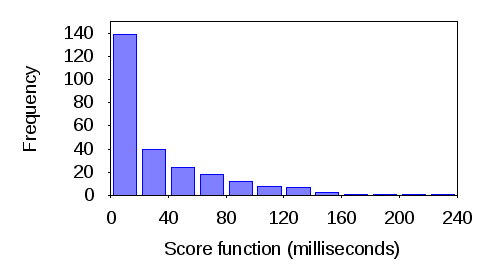}%
	}%
	\caption{Histogram of score function ($\chi$) values.}
	\label{fig:delta_histogram}
\end{figure}

\remove{
We emphasize that we can use this technique to measure the scoring
function and $\Delta$ for any arbitrary workload, and not just for the
synthetic workloads used in our experiment. We can also choose any key
for our measurements, and do not need to introduce additional
operations into the workload in order to obtain
measurements. Moreover, our methods are minimally intrusive, in that
we don't need to inject load into the system and disturb the original
workload to measure the workload.
}

We computed $\chi$ and $\Delta$ from collected timing information, as described in Section~\ref{sec:pos}.
Figure~\ref{fig:delta_histogram} is a histogram of positive $\chi$
values for all keys.  Each point represents the relative staleness
observed by some read operation on some key.  The value of $\chi$ ranges
from 1ms to 233ms, and the margin of error due to clock skew is around 1ms.
In comparison, Wada et al.\ report much higher maximum staleness levels in their experiments using Amazon's SimpleDB
(see Figures 2 and 3 in \cite{wada:cidr}).

\remove{
The histogram in Figure \ref{fig:delta_histogram} depicts the
distribution of $\chi$ values for one key in one run.  Each
point represents the relative staleness observed by a read operation
in a cluster (e.g., subset of Cassandra nodes). The maximum point
equals the $\Delta$ value for that combination of execution and
key. The range [1ms-4ms] is most interesting, and seems to follow
(approximately) an exponential distribution. This distribution
indicates that although a system like Cassandra only promises eventual
consistency, the \emph{observed} data staleness is less that 4ms most
of the time. The scoring function and $\Delta$ values can quantify the
consistency properties of a system at run-time, which can be
invaluable, for example, to clients monitoring the service level
agreements with the storage provider and the client.
}

\remove{We repeat the above experiment $10$ times and obtain the $\Delta$ value
for each run. The $\Delta$ values have an average of 53ms with a
standard deviation of 12ms, indicating a low degree of staleness in the worst case.
We expect to observe larger $\Delta$'s for heavier workloads, and much larger
  $\Delta$'s still for executions containing failures.}

\begin{figure}[tbp]
    \centering
    \vspace{-10pt}
    \subfigure{
      \includegraphics[width=0.93\linewidth]{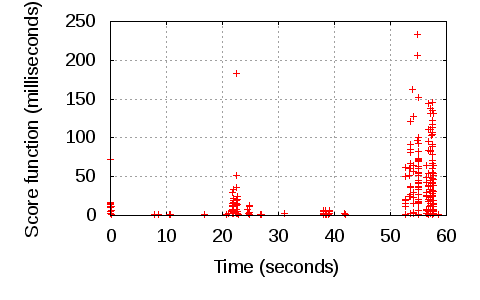}%
    }
    \caption{Time series plot of score function ($\chi$) values.}
    \label{fig:time_series}
\end{figure}

Figure~\ref{fig:time_series} shows a time series plot of the $\chi$ values for all keys).
This visualization allows us to observe how staleness varies over time, in contrast to the
  distribution of staleness values captured in Figure~\ref{fig:delta_histogram}.
In Figure~\ref{fig:time_series}, the x-axis depicts the approximate time when a read returns a stale value, and the y-axis depicts the corresponding $\chi$ value. 
Most of the data points are concentrated near the x-axis, as we expect based on the histogram,
  and furthermore there are a few visible ``inconsistency spikes''. 

Finally, we measured the overhead of instrumentation that is required to compute the staleness metric and observed a performance loss of less than five percent with instrumentation enabled.



\remove{
This visualization allows us to observe how staleness
varies over time (for a history of operations), instead of only the
worst-case staleness captured in Figure~\ref{fig:delta_histogram}. We
observe a number of spikes in the plot, which may indicate instances
of high contention, and failure patterns in the system. We will
explore this line of work in future.
}

\section{Conclusions and future work}

In this paper, we present a client-centric benchmarking methodology
for understanding eventual consistency in distributed key-value
storage systems.  Our methodology measures observed, rather than
worst-case, consistency.  It extends the popular YCSB benchmark to
measure the staleness of data returned by reads using the concept of
$\Delta$-atomicity \cite{gls:fun}.  Because our technique does not
inject operations into the workload, it measures consistency in a more
faithful manner than prior benchmarks.  By measuring consistency in a
system-agnostic manner, we provide a quantitative methodology for
examining the performance vs.\ consistency trade-offs across various
key-value system architectures.


Using a preliminary implementation of our methodology, we demonstrate that the staleness of data in
Cassandra exhibits a long and thin tail.
That is, the worst-case staleness is much higher than the typical staleness
of data returned by read operations.  This observation has implications for a system
administrator when deciding how to configure or deploy a system like
Cassandra---depending on the desired performance and deployment size,
the choice of replication factor and quorum sizes can be guided by
our benchmark results rather than guesswork.

We are actively extending our work to consider runs with failures.
Events such as network partitions, software crashes, or
device failures may trigger special execution paths in the system and
result in different consistency behaviors.
Our goal in future work is to stage experiments involving such failures
  through additional modifications to the $\Delta$-enabled YCSB suite.

\vspace{-2.5mm}
\paragraph{Acknowledgments}
Thanks to Steve Uurtamo for conducting preliminary experiments.
We are also grateful to the anonymous reviewers for their helpful feedback.
\newpage
\vspace{-2.5mm}
{\footnotesize \bibliographystyle{acm}
\bibliography{bibfile}}


\end{document}